\documentclass[aps,prb]{revtex4-1}
\usepackage{amsmath}
\usepackage{amsfonts}
\usepackage{amssymb}
\usepackage{amsthm}
\usepackage{graphicx}
\usepackage{bbm}
\usepackage{bm}
\usepackage{bbold}
\usepackage{physics}
\newcommand{\sinc}{{\rm sinc}}

\newcommand{\stau}{\tau}
\newcommand{\phins}{\phi_{n\stau}}
\newcommand{\stauzt}{\tau_z(t)}
\newcommand{\bstns}{\expval{\stauzt}_{n\stau}}
\newcommand{\ketn}{\ket{n}}
\newcommand{\bpsn}{b^{\stau}_n}
\newcommand{\bmsn}{b^{-\stau}_n}

\newcommand{\Dn}{\Delta_n}
\newcommand{\sssp}{\sum_{ss^\prime}}
\newcommand{\smmp}{\sum_{mm^\prime}}

\begin{document}

%----------------------------------------------------------------%
\title{Photoinduced polarization enhancement in biased bilayer graphene in the  Landau level regime}
\author{Alexander L\'{o}pez}
\email[To whom correspondence should be addressed. Electronic
address: ]{alexlop@espol.edu.ec}
\affiliation{Escuela Superior Polit\'ecnica del Litoral, ESPOL, Departamento de F\'isica, Facultad de Ciencias Naturales y Matem\'aticas, Campus Gustavo Galindo
 Km. 30.5 Via Perimetral, P. O. Box 09-01-5863, Guayaquil, Ecuador} 
 \author{Bertrand Berche}
 \affiliation{Laboratoire de Physique et Chimie Th\'eoriques,\\
 UMR Universit\'e de Lorraine - CNRS 7019,\\  
	 B.P. 70239, F - 54506 Vand{\oe}uvre l\`es Nancy Cedex, France, EU}
\author{John Schliemann}
 \affiliation{ Institute for Theoretical Physics, University of
Regensburg, D-93040 Regensburg, Germany}
\author{Francisco Mireles}	
\affiliation{Departamento de F\'isica, Centro de Nanociencias y Nanotecnolog\'ia, Universidad Nacional Aut\'onoma 
 de M\'exico, Apdo. Postal 14, 22800 Ensenada B.C., M\'exico}
\author{Benjamin Santos}
 \affiliation{INRS-EMT, Universit\'e du Qu\'ebec, 1650 Lionel-Boulet, Varennes, Qu\'ebec J3X 1S2, Canada }
%\date{\today}

\begin{abstract}
We investigate the charge carrier dynamics in bilayer graphene subject to monochromatic laser
irradiation within the Landau level quantization regime. Even though the radiation field does not lift the energy degeneracy of the lowest
Landau levels (LLs), it nevertheless has a strong effect on the photoinduced pseudospin polarization response for higher LLs
($n\ge2$). Our results show that the photoinduced bandgaps lead to a finite response of the averaged pseudospin polarization with nontrivial
oscillating behavior. It is shown that the contribution from these higher LL transitions turns out to be crucial to achieve an enhanced
photoinduced polarization in radiated bilayer graphene. The experimental feasibility of our findings is also discussed.
\end{abstract}

\maketitle
%----------------------------------------------------------------%
\section{Introduction}
The proposals of driven phase transitions\cite{oka} in graphene\cite{novoselov1,geim,guinearmp,kane}, so called Floquet topological insulators\cite{fti}, 
opened the road to explore many interesting phenomena in the family of two-dimensional materials that support Dirac fermions when they are subject to 
periodically modulated time-dependent interactions\cite{foa,ganichev,topological1,topological2,auerbach,rabill}. These photoinduced topological phases extended
the static results of topological insulators to the dynamical regime showing that these topological phases could be dynamically generated even if the material
showed a trivial topological phase in the static scenario. Indeed, within the static regime, both experimental and theoretical works have shown that 
the transport properties of topological insulating materials present very distinct  properties that contrast those of conventional two dimensional electron 
gases 2DEG. 
An interesting example of those distinct features occurs in the Landau level structure of monolayer samples of graphene which, in contrast to the semiconductor 2DEG, 
shows a nonequidistant energy spectrum that, in turn, could allow the realization of a tunable laser in the Terahertz domain\cite{rusin,aoki,wendler}.\\ 
Moreover, the lowest LL in monolayer graphene can only be occupied in one 
of its sublattice degrees of freedom. This special behavior of the $n=0$ Landau level (LL) in 
graphene monolayer renders the associated quantum Hall effect particularly interesting since at the charge neutral Dirac point it splits into four 
sublevels\citep{qhe-new} at high magnetic fields (one for each valle $K$, $K'$ and one per spin state. When considering the bilayer graphene 
scenario\cite{bilayer,bilayer2,bilayer3,bilayer4,bilayer5,bilayer6,mireles} at low energies, the effective charge carriers behave as massive chiral particles in absence of a quantizing magnetic field. 
In addition, it has been experimentally shown\cite{oostinga} that it is possible to realize a gate induced insulating phase in bilayer graphene. Upon addition of a perpendicular magnetic field, the low energy excitations show an energy degeneracy at $n=0,1$ which is manifested in the transport measurements as additional van Hove singularities in the density of states\cite{bilayer}. Another experimental work shows that for the bilayer case in presence of a quantizing magnetic field, a $2\pi$ Berry phase is observed\cite{bexp} which contrasts the $\pi$ Berry phase acquired by Dirac fermions in monolayer graphene. One could expect that upon introduction of electromagnetic radiation, novel features should be feasible to be realized  in the dynamical evolution of the 
effective charge carriers at low energies. Indeed these novel features have already been explored in irradiated bilayer samples with and without trigonal warping effects\citep{bilayer,trigonal,biased1,biased2}. However, the interplay of an applied bias voltage and circularly polarized monochromatic radiation on the pseudospin polarization  
of bilayer graphene and to what extend enhanced polarization inversion capabilities can be achieved is still a physics to be explored in bilayer graphene and is one of the main focus of the present work.
 
\noindent The standard theoretical approach to describe the electronic properties of these materials rely on either first principles (DFT) numerical 
calculations or the use of a tight binding description, either of these gives 
detailed account of a number of the electronic features of both monolayer as well as bilayer samples\cite{polini}. However, 
it is well known that many phenomena of interest emerge already from the low energy physics\cite{rpp}. In such regime, one can find analytically tractable 
models which, in turn, can shed light on the underlying physical processeses allowing for instance, the realization of 
novel transport features. In particular, within this low energy regime, most of the relevant physical features of bilayer graphene can be captured via an effective two-band model.\\
\noindent  In this work we theoretically analyze the dynamical manipulation of the LL structure of spinless charge carriers in biased bilayer graphene (in the Bernal stacking 
configuration), subject to  a periodically driving radiation field applied perpendicular to the sample. In our approach we make use of Floquet's theorem\cite{milena,chu,kohler,platero,shirley,sambe} 
to recast the dynamics in an explicitly time-independent fashion but without the need to resort to the brute force numerical solution of the infinite-dimensional 
Fourier-mode expansion technique. Our approach has the advantage 
of providing an analytical description of the driven evolution of relevant physical quantities such as the pseudospin polarization. Since we are mostly interested in the leading order dynamical effects 
induced by the radiation field, we shall neglect trigonal warping effects that render the energy spectrum anisotropic at very low energies ($<1\textrm{meV}$)\cite{trigonal}. 
We will also discard any spin-orbit effects (see, for instance, reference \cite{mireles} 
for the interplay of spin-orbit effects and quantizing magnetic fields). In doing so, our analysis allows us to explicitly address each LL in an independent fashion and we show that the photoinduced
bandgap depends on the Landau level index such that  the $n\ge2$ LL quasienergy spectrum 
gives rise to a level-dependent bandgap. We then use this in order to infer the physical consequences in the dynamical evolution of physical quantities.\\ 
Although bilayer graphene can also be experimentally realized in the AA stacking, where the two layers are laid on top of each other with the corresponding A2 (B2) atoms laying on top of the A1 (B1) atoms, this configuration shows a static unbiased energy spectrum consisting of two shifted copies of the monolayer graphene spectrum\cite{37,38}, with shifted Dirac cones separated by an energy of the order of the interlayer coupling. Upon introduction of a biased potential U among the layers, a static bandgap develops. Therefore, although the interlayer coupling is larger than the typical values of the bias gate voltage, the dominant physical mechanism for static bandgap generation is the bias voltage term. Indeed, in the absence of gate voltage the Landau level physics under electromagnetic radiation would be that of two decoupled copies of monolayer graphene which we have already addressed in reference\cite{39}. This is the reason why we are choosing the AB configuration in order to describe a distinct physical scenario as that of monolayer graphene. 

Thus, it is shown that in contrast to the single layer scenario, in the AB stacking configuration, the additional layer degree of freedom in bilayer graphene offers a richer physical structure for the pseudospin polarization. We find that its amplitude and decaying time can be enhanced via the radiation field, within experimentally accessible parameter regimes. In addition, the quasienergy spectrum LL anti-crossings emerging under radiation offer means to explicitly address an effective two-level system dynamics that is independent of the intensity of the quantizing magnetic field. We show that by properly tuning the laser parameters the photoinduced bandgap of the different Landau levels can lead to regimes from semi-conducting to metallic transitions with a finite to vanishing effective bandgap transition, respectively. We find that the photoinduced bandgap opening behavior enables a larger effective photoinduced polarization contribution for the higher order ($n\ge2$) LL states. As we will show below, our results on the photoinduced enhancement of the polarization effects could also lead to potential applications in quantum optics. In this realm, we could suggest using irradiated bilayer graphene in quantum optics for realizing a tunable laser taking advantage of the tunable effective bandgap that we have obtained.

\noindent The paper is organized as follows. In section II we present the model and summarize the results for the quasienergy spectrum  and the dynamics of the 
pseudospin polarization. Next, in section III we discuss the main results and we give concluding remarks, arguing on the possible experimental implementation of our 
proposed theoretical setup. Finally, in the appendix we summarized some 
mathematical calculations used during the perturbative analysis.
\section{Model}\label{sec1}
\begin{center}
\begin{figure}
\includegraphics[height=5cm]{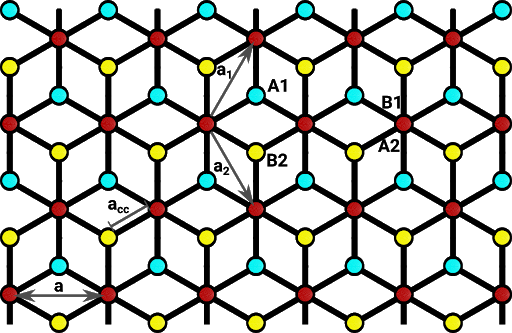}\hspace{0.25cm}\includegraphics[height=6cm]{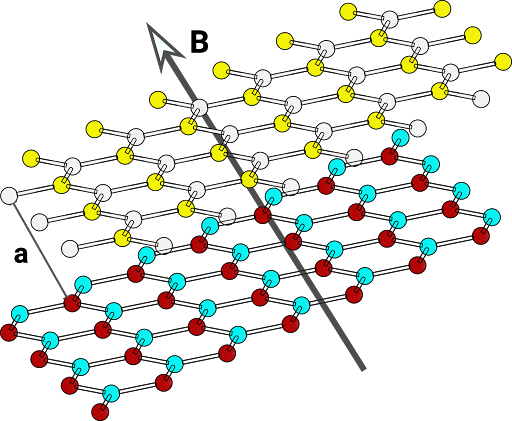}  
\caption{(Color online) Bernal stacking configuration with dimer states $B_1-A_2$ (red sites) and non-dimer sites $B_2$ (yellow sites) and $A_1$ (blue sites).}
\end{figure}
\end{center}
Let us begin by considering the Landau level structure of biased  bilayer graphene subject  
to intense circularly polarized monochromatic  radiation. For definitness, we consider the planes of the layers to be the $x-y$ plane, and the perpendicular direction
to be that of the quantizing static magnetic field. We measure the momentum of the particles from the K point. Thus, the presence of the magnetic field modifies the momentum 
operator as $\vec{p}\rightarrow \vec{p}+e\vec{A}\equiv \bf{\pi}$, where the vector potential satisfies $\nabla\times \vec{A}=\vec{B}$ and we are considering 
$\vec{B}=B \hat{z}$. Our model for the bilayer graphene is such that the layers are arranged according to Bernal $A_2-B_1$ stacking (i.e. atom $A_2$ from the 
upper layer lies directly on top of atom $B_1$ in the lower layer, see figure FIG. 1).  In this case,  near the $K$ Dirac point at low energies and taking into 
account the time-dependent  radiation field, we have a $4\times4$ Hamiltonian $\mathcal{H}(t)=\mathcal{H}_0+\mathcal{V}(t)$, where $\mathcal{H}_0$ is the 
static part and $\mathcal{V}(t)$ describes the light-matter interaction. The static contribution is explicitly given by ($\hbar=1$)
\begin{equation}\label{bilayer}
\mathcal{H}_0=\left(
\begin{array}{cccc}
 -U&\omega_c\hat{a}&0&0\\
 \omega_c\hat{a}^\dagger& -U&\gamma&0\\
0&\gamma&U&\omega_c\hat{a}\\
0&0&\omega_c\hat{a}^\dagger&U
\end{array}
\right),
\end{equation}
where we have introduced the operators $\hat{a}=(\pi_x-i\pi_y)/\omega_c$ and $\hat{a}^\dagger=(\pi_x+i\pi_y)/\omega_c$  where the quantity 
$\omega_c=\sqrt{2}v/\ell_B$ describes the cyclotron frequency for massless Dirac fermions in monolayer graphene, $\gamma$ is the interlayer coupling and $U>0$ a 
biasing gate voltage strength. In addition, $v\approx10^6$ m/s is the Fermi velocity of charged particles in monolayer graphene, whereas $\ell_B^{-2}=eB$ 
is the magnetic length defined in terms of the strength of the quantizing magnetic field $B$ and $e$ is the electric charge. 
Notice that the static Hamiltonian $\mathcal{H}_0$ is written in the basis such that its upper left and lower right $2\times2$ sub-block matrices describe the 
lower and upper layers with inequivalent atoms labelled as $A_1, B_1$ and $A_2, B_2$, respectively. In addition, the upper right and lower left $2\times2$ 
sub-block matrices describe the interlayer coupling of strength $\gamma$ (between $A_1-B_2$ sites). Here we are neglecting warping effects caused  by  
weaker couplings $\gamma_3$ among sites  $A_1-B_2$ as well as a $\gamma_4$ coupling among $A_1-A_2$ and $B_1-B_2$ sites which give rise to electron-hole 
assymetry. Additional hoppings can be neglected in a zero order bilayer Hamiltonian. It is known that they can play a role at very low energies 
$<1\,\textrm{meV}$. 
The low energy physics is described via an effective two-band Hamiltonian to leading order in the interlayer coupling. In what follows, we shall concentrate in the range of energies $|\varepsilon|\ll\gamma$, so in such case the reduced
 Hamiltonian takes the form, 
\begin{equation}\label{h20}
H_{2}=\left(
\begin{array}{cc}
 U&\Omega_c(\hat{a}^\dagger)^2\\
\Omega_c(\hat{a})^2& -U
\end{array}
\right),
\end{equation}  
where $\Omega_c=\omega_c^2/\gamma=2v_F^2eB/\gamma$. In this manner, the largest energy scale is determined by the interlayer coupling parameter $\gamma$. 
Within this regime, the associated Landau spectrum in the absence of radiation reads $E_{ns}=sE_n$ with $E_n=\sqrt{U^2+n(n-1)\Omega_c^2}$, for $n\neq0,1$. The corresponding eigenstates are explicitly given by
\begin{equation}\label{evec0}
|\phi_{ns}\rangle=\left(
\begin{array}{c}
 b^{s}_ n|n\rangle\\
  sb^{-s}_{n}|n-2\rangle
\end{array}
\right),
\end{equation}
where  $n= 2,3\dots$ label the unperturbed Landau levels and $s=\pm1$ label the conduction and valence band. We have also introduced the coefficients 
\begin{equation}\label{cs}
b^s_{n}=\sqrt{\frac{E_n+sU}{2E_n}}.
\end{equation}
The degenerate LL corresponding to $n=0,1$ have energy $E_{0,1}=U$ and read
\begin{equation}\label{01}
|\phi_{0}\rangle=\left(
\begin{array}{c}
 |0\rangle\\
  0
\end{array}
\right),\quad
|\phi_{1}\rangle=\left(
\begin{array}{c}
 |1\rangle\\
  0
\end{array}
\right).
\end{equation}

We notice that in the presence of the bias voltage the electron density of the states can be changed and the associated bandgap can be tuned accordingly\cite{biased1,biased2}. 
Yet, we will assume a small bias such that the electron density is essentially fixed. Thus, our results will be valid within the low light-matter coupling strength. 
Using this approximation, one can for instance evaluate the effective effective inter Landau level polarization
polarization $\langle\tau_z\rangle$, where $\tau_z$ is a $2\times2$ Pauli matrix. 
Thus, one neads to evaluate $\langle\tau_z\rangle_{ns}=\langle\phi_{ns}|\tau_z|\phi_{ns}\rangle$. 
After some algebraic manipulation we get
\begin{equation}\label{polar0}
\langle\tau_z\rangle_{ns}=\frac{sU}{E_n},
\end{equation} 
which certainly implies that biasing the system does indeed introduces a finite pseudospin polarization via the energy bandgap and this manifest as a finite value of the effective inter Landau level polarization $\langle\tau_z\rangle_{ns}$.\\

\noindent We now deal with the photoinduced effects. In order to take into account the light-matter interaction in the model, we start from the standard minimal coupling interaction term $-e{\bf v}\cdot{\bf A}(t)$, which can be introduced, via the Peierls 
substitution, in the full $4\times4$ Hamiltonian. 
Hence, at the Dirac point, the effects of the driving field is described in the basis of ec. (\ref{bilayer}) by 
\begin{equation}
\mathcal{V}(t)=e v_F\mathbb{1}\otimes\mathbb{\sigma}\cdot{\bf A}(t),
\end{equation} 
with $\mathbb{1}$ being the $2\times2$ unit matrix, $\mathbf{\sigma}=(\sigma_x,\sigma_y)$ a vector of Pauli matrices and 
${\bf A}(t)=A(\cos\omega t,\sin\omega t)$ the in-plane associated radiation field which is related to the electic field via 
$\bm{\mathcal{E}}(t)=-\partial_t {\bf A}(t)$, where $A=\mathcal{E}/\omega$, with $\mathcal{E}$  and $\omega=2\pi/T$ being respectively, the amplitude and frequency of the 
radiation field, with $T$ being its period. 
This periodic interaction makes the total Hamiltonian 
\begin{equation}\label{time}
\mathcal{H}(t)=\mathcal{H}_0+\mathcal{V}(t),
\end{equation}
periodic in time $\mathcal{H}(t+T)=\mathcal{H}(t)$, with $T=2\pi/\omega$ the period of oscillation of the driving field. 
Then, the time-dependent contribution reads
\begin{equation}\label{bilayerint}
\mathcal{V}(t)=\xi\left(
\begin{array}{cccc}
 0&e^{-i\omega t}&0&0\\
 e^{i\omega t}& 0&0&0\\
0&0&0&e^{-i\omega t}\\
0&0&e^{i\omega t}&0
\end{array}
\right).
\end{equation} 
where we have introduced the effective light-matter coupling strength $\xi=e\mathcal{E}v_F/\omega$, given in terms of 
$\mathcal{E}$ and $\omega$ the amplitude and frequency of the electric field, respectively. 
\noindent We assume that the beam radiation spot is large enough compared to the lattice spacing so we can ignore any spatial variation.
On the other hand, we notice that the static Hamiltonian (\ref{bilayer}) commutes with the operator
\begin{equation}
\mathcal{N}_a=\left(
\begin{array}{cc}
 N_a& 0\\
 0& N_a-\mathbb{1} 
\end{array}
\right),
\end{equation}
where the operator $N_a$ is defined as
\begin{equation}\label{nz}
N_a=\Big(a^\dagger a+\frac{1}{2}\Big)\mathbbm{1}+\frac{\sigma_z}{2}.
\end{equation} 
\noindent 
We can then perform a time-dependent unitary transformation $\mathcal{H}_F=\mathcal{P}^\dagger(t)\mathcal{H}(t)\mathcal{P}(t)-i\mathcal{P}^\dagger(t)\partial_t\mathcal{P}(t)$ where $\mathcal{P}(t)=e^{-i\mathcal{N}_a\omega t}$, given explicitly as  
\begin{equation}\label{pp}
\mathcal{P}(t)=\left(
\begin{array}{cc}
 e^{-iN_a\omega t}& 0\\
 0& e^{-i(N_a-\mathbb{1})\omega t} 
\end{array}
\right)
\end{equation}
which yields the time-independent Floquet Hamiltonian\cite{milena,chu}
\begin{equation}\label{floquetbi2}
\mathcal{H}_F=U\tau_z\otimes\mathbb{1}+\left(
\begin{array}{cc}
H_F&\gamma\sigma_-\\
\gamma\sigma_+& H_F+\mathbb{1}\omega
\end{array}
\right),
\end{equation} 
where $H_F$ is given by 
\begin{equation}\label{heff}
H_{F}=\omega_c(a^\dagger\sigma_-+a\sigma_+)-N_a\omega+\xi\sigma_x.
\end{equation}
We focus our analysis in the far infrared frequency domain\cite{ganichev}, where the laser energy is of the order of 
$\omega\approx 10\, \textrm{meV}$ and the effective radiation field intensity has the value $\mathcal{E}\sim 1\, \textrm{kV/m}$, for which $\xi\approx 10\mu \textrm{eV}$. Yet, we will show that our results 
could apply for larger electric field intensities $\mathcal{E}\sim 0.15\, \textrm{MV/m}$ for which one gets for the coupling constant 
$\xi\approx 10\, \textrm{meV}$. For frequencies $\omega$ in the Terahertz ($\omega=3$ THz) one gets  $\xi\approx\omega$. 
This is an order of magnitude smaller than the Landau level separation $\omega_c\approx 116\, \textrm{meV}$, for $B= 10\, \textrm{T}$, which is a typical 
experimental value at such fields. 
For larger frequencies and stronger magnetic field intensities, the ratio $\xi/\omega_c$ tends to be smaller and our approximation scheme should provide 
values for the physical quantities that could be closer to those experimentally achievable. 
Thus, we can perform a perturbative treatment in the effective coupling parameter $\lambda=\xi/\omega_c\ll1$. We also notice that for 
intermediate values of the quantizing magnetic field satisfying $\xi\ll\omega_c,\gamma$ we can write down an effective two-band Hamiltonian among 
non-dimer sites; this is justified by recalling that quasienergies can be defined within the first Brillouin zone $-\omega/2<\varepsilon<\omega/2$ and thus 
the effective two-band Hamiltonian approximation can be justified whenever $\omega\ll\gamma$ (see discussion below).\\ 
\noindent Let us now write down a perturbative Hamiltonian using the perturbation parameter of interest $\lambda=\xi/\omega_c$. 
For this purpose we use the antihermitian operator $I_-=a^\dagger\sigma_--a\sigma_+$, and build the $4\times4$ unitary matrix
\begin{equation}\label{dia4}
\mathcal{T}=\left(
\begin{array}{cc}
e^{-\lambda/2 I_-}&0\\
0& e^{-\lambda/2 I_-}
\end{array}
\right),
\end{equation} 
which transforms the Floquet Hamiltonian given in equation (\ref{floquetbi2}) as $\mathcal{H}_F\rightarrow \tilde{\mathcal{H}}=\mathcal{T}^\dagger \mathcal{H}_F\mathcal{T}$.
Since $\lambda$ is small we restrict our analysis up to first order in the effective perturbation parameter 
$\lambda$. Introducing the shifted harmonic oscillator operators $b=a+\lambda$, we get (to leading order in $\lambda$) the effective Floquet Hamiltonian
\begin{equation}\label{bifin}
\tilde{\mathcal{H}}\approx U\tau_z\otimes\mathbb{1}+\left(
\begin{array}{cc}
H&\gamma\sigma_-\\
\gamma\sigma_+& H+\mathbb{1}\omega
\end{array}
\right),
\end{equation} 
with $H$ given by 
\begin{equation}\label{hfin}
H=\omega_c\Big(b^\dagger\sigma_-+b\sigma_+\Big)-\omega N_b -\xi N_b\sigma_z,
\end{equation} 
that takes into account corrections of order $\xi$. To get the result (\ref{bifin}) we have neglected the off-diagonal contributions
\begin{equation}\label{neglect1}
\Delta\mathcal{V}_{OD}=\frac{\lambda\gamma}{2}\left(
\begin{array}{cc}
0&b\sigma_z\\
b^\dagger\sigma_z& 0
\end{array}
\right)-\frac{\lambda^2\gamma}{2}\left(
\begin{array}{cc}
0&\sigma_z\\
\sigma_z& 0
\end{array}
\right),
\end{equation} 
that can be treated by nondegenerate perturbation theory.  They give corrections of order $O(\lambda^2\gamma^2)\approx\xi^2$ and $O(\lambda^4\gamma^2)\approx\xi^4$, respectively. Then,  
they turn out to be less important than the last coupling term given in (\ref{hfin}) that give corrections of order $\xi$.  In addition, we have also neglected the higher order
diagonal terms
\begin{figure*}\label{fig2}
\includegraphics[height=6cm]{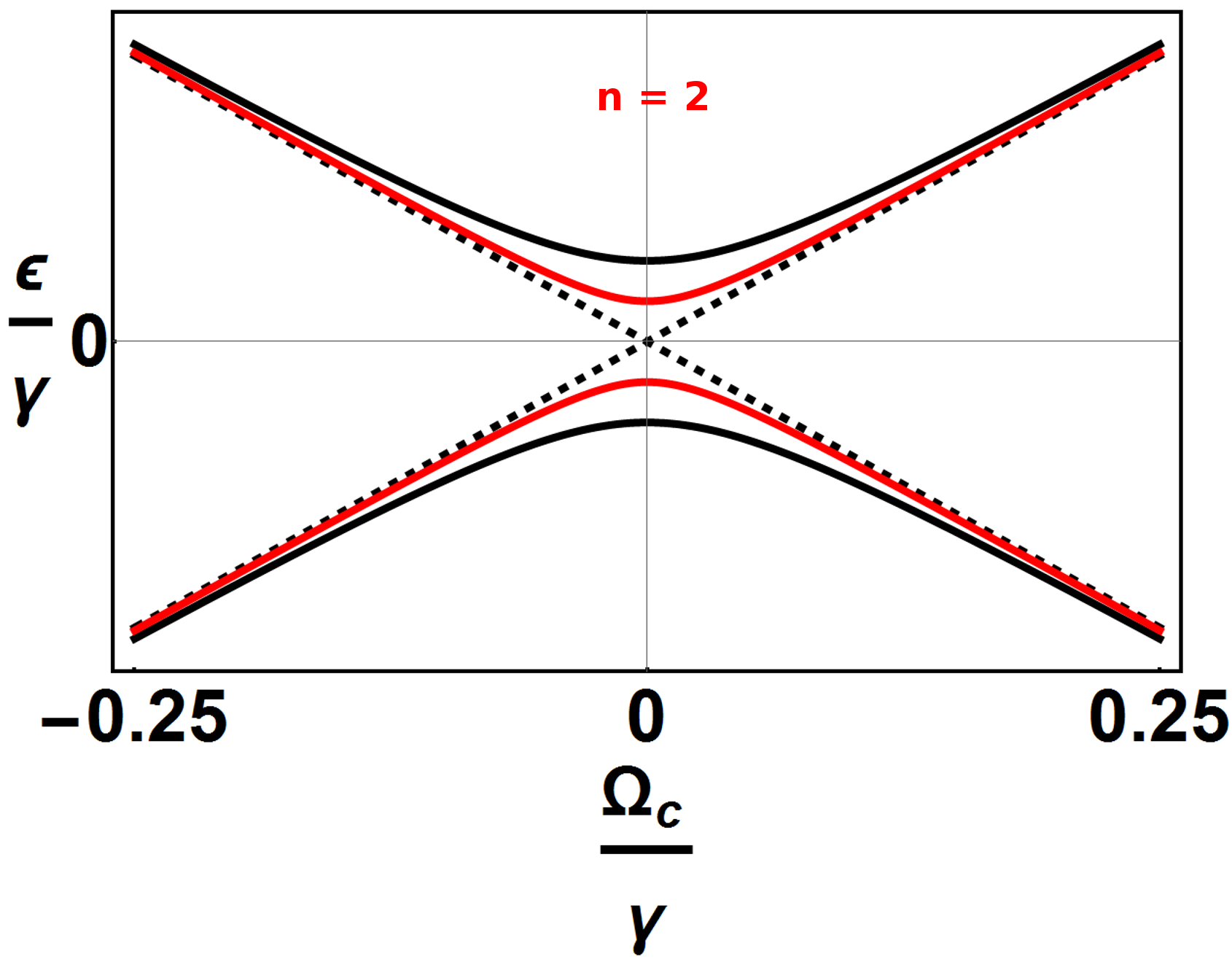}\hspace{.5cm}\includegraphics[height=6cm]{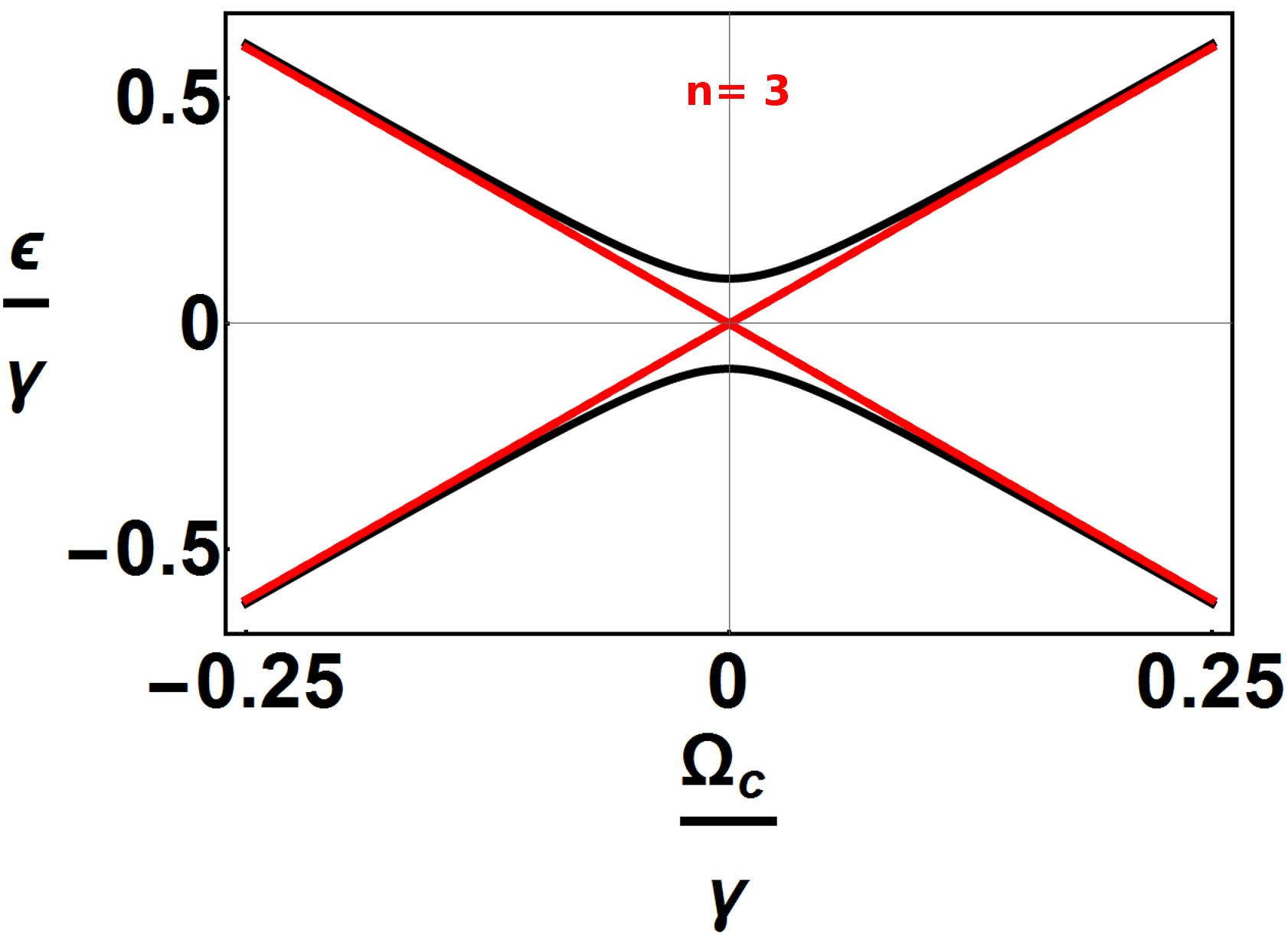}\\
\includegraphics[height=6.cm]{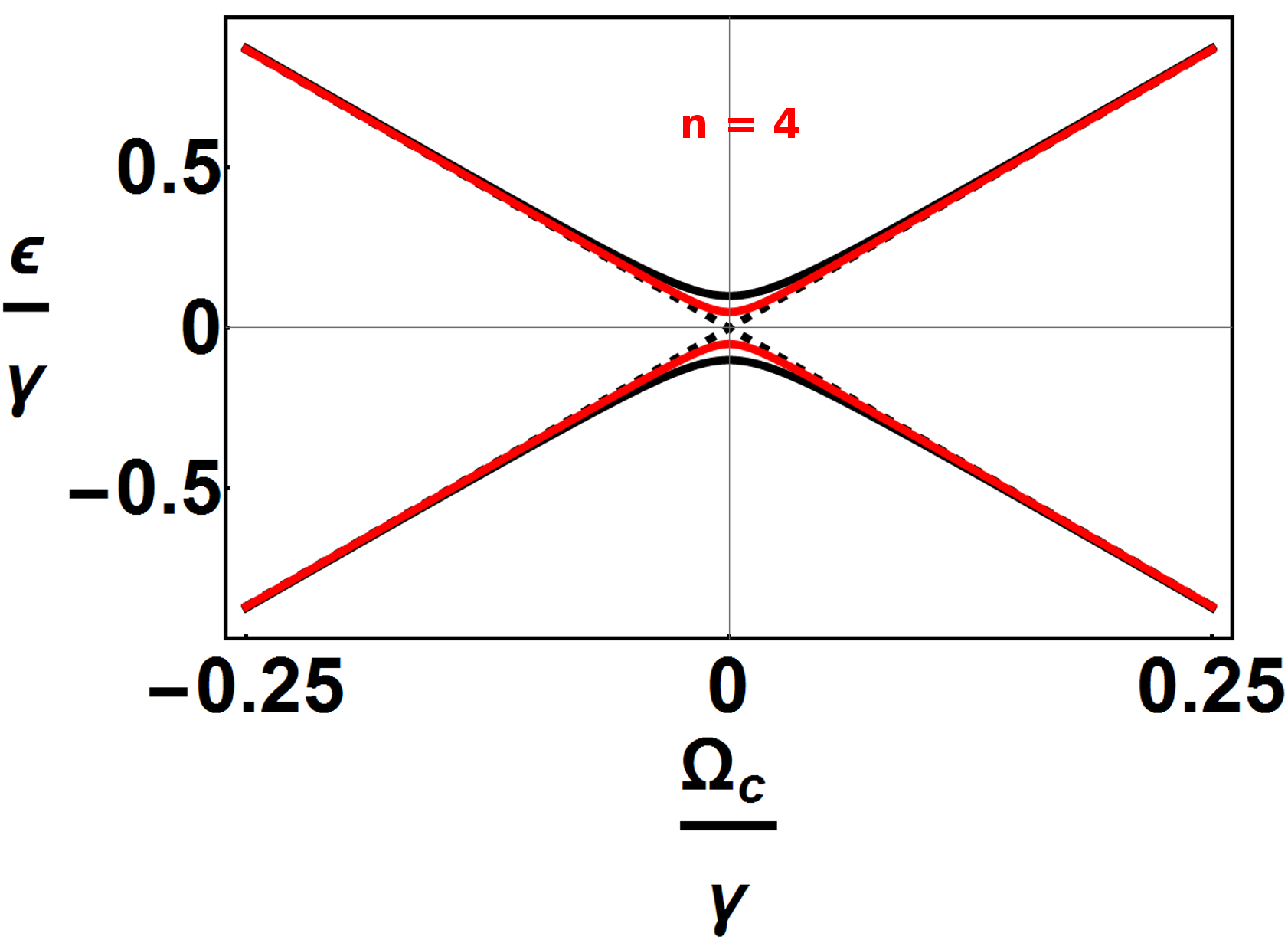}\hspace{.5cm}\includegraphics[height=6.cm]{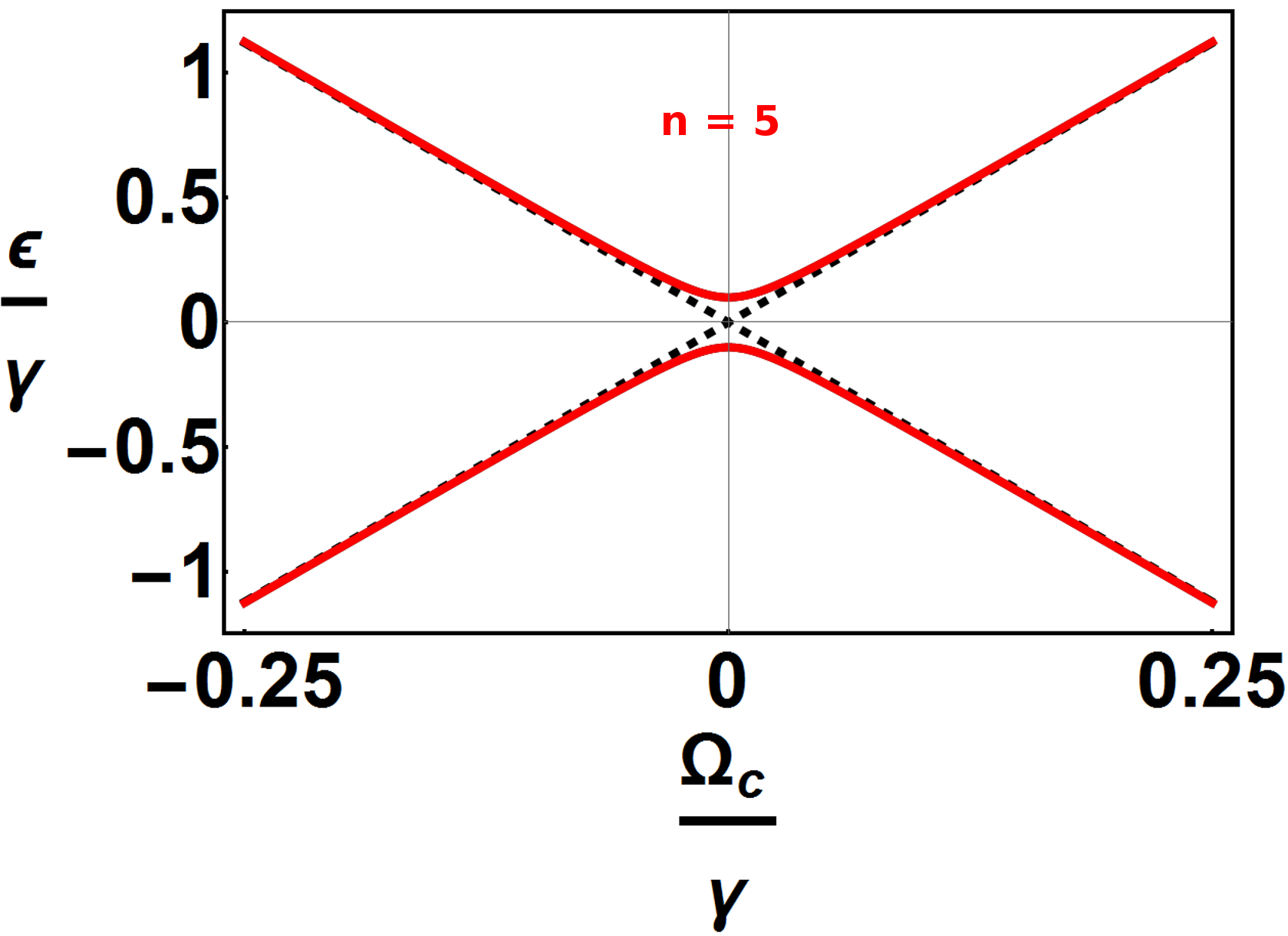}
\caption{(Color online) Landau level spectrum for $n=2$ (upper left), $n=3$ (upper right), $n=4$ (lower left), $n=5$ 
(lower right) as functions of normalized cyclotron energy. The dashed (continuous) black lines represents the static unbiased (biased) 
spectrum whereas the continuous red (light) line corresponds to the biased driven effective low energy spectra. We have set an effective bias 
$U/\gamma=0.1$ whereas the effective coupling is set to $\xi/\gamma=0.05$. For the chosen set of parameters we can get for $n=3$ a biased driven spectrum that mimics the static unbiased regime and  for $n=5$ a driven biased spectrum
indistinguishable from the undriven biased regime (see discussion in the main text).}
\end{figure*}
\begin{equation}\label{neglect2}
\Delta\mathcal{V}_{D}=\lambda\omega(b+b^\dagger-\lambda)\left(
\begin{array}{cc}
\mathbb{1}&0\\
0&\mathbb{1} 
\end{array}
\right)+\frac{\lambda\xi}{2}(b+b^\dagger-2\lambda)\left(
\begin{array}{cc}
\sigma_z&0\\
0&\sigma_z
\end{array}
\right),
\end{equation}
that could also be dealt by higher order perturbation theory.
The contributions 
(\ref{neglect1}) and (\ref{neglect2}) can be relevant in the regime of small 
quantizing static magnetic fields $B\approx 1T$, but can be discarded in our following discussion since, a posteriori shows that these higher order 
contributions do not qualitatively change our main results.\\

\noindent For bilayer graphene some works have reported 
values for $\gamma\approx\,400\textrm{meV}$ (see \cite{rpp} and references therein). Then, for a quantizing magnetic field used in experimental setups\cite{ganichev} $B\sim 10\,\textrm{T}$, such that $\Omega_c\approx 0.25\gamma$ we can safely use this effective low energy two-band approximation. 
Thus, for our purposes we can work within the effective two-band reduced Floquet Hamiltonian (\ref{bifin}). 
First, we find useful to perform a unitary transformation $\mathcal{H}_{B}=\tilde{\mathcal{R} \mathcal{H}}\mathcal{R}^{-1}$ where $\mathcal{R}$ explictly reads
\begin{equation}
\mathcal{R}=\left(
\begin{array}{cccc}
 0&0&0&1\\
 1& 0&0&0\\
0&1&0&0\\
0&0&1&0
\end{array}
\right),
\end{equation}
that leads to,
\begin{widetext}
\begin{equation*}\label{4x4}
\mathcal{H}_B+\omega\mathcal{N}_b=\left(
\begin{array}{cccc}
 U-(n_b-1)\xi&0&0&\omega_cb^\dagger\\
 0& -U+(n_b+1)\xi&\omega_cb&0\\
0&\omega_cb^\dagger&U+n_b\xi&\gamma\\
\omega_cb&0&\gamma&-U-n_b\xi
\end{array}
\right),
\end{equation*}
\end{widetext}
 where $\hat{n}_b=\hat{b}^\dagger\hat{b}$. The corresponding Floquet eigenstate is $\mathcal{R}|\Phi\rangle=|\Psi\rangle$, which has the  two-component bi-spinor form $|\Psi\rangle\rangle=(|\psi_l\rangle\quad |\psi_h\rangle)^T$, where the upperscript $T$ denotes transpose and we have separated the 
 lower energy $|\psi_l\rangle$ and $|\psi_h\rangle$ spinors corresponding to non-dimer and dimer coupling among the two layers. Thus, after eliminating the 
 high energy spinor component we get the effective low energy two-band quasienergy problem $H_{2F}|\psi\rangle=\varepsilon|\psi\rangle$, where the effective two-band Floquet Hamiltonian reads now
 \begin{equation}\label{low-high}
H_{2F}=\left(
\begin{array}{cc}
 U-(\hat{n}_b-1)(\xi+\omega)&\Omega_c(\hat{b}^\dagger)^2\\
\Omega_c(\hat{b})^2& -U+(\hat{n}_b+1)(\xi-\omega)
\end{array}
\right),
\end{equation}  
with $\Omega_c=2v_F^2eB/\gamma$. The effective Hamiltonian (\ref{low-high}) is valid whenever the condition $\gamma\gg \Omega_c\,, U\,,\xi$ is fulfilled. 
The Hamiltonian given in equation (\ref{low-high}) has quasienergies 
\begin{equation}
\epsilon_{ns}=s\sqrt{[U-(n-1)\xi]^2+\Omega_c^2n(n-1)}=s\epsilon_n,\quad \text{mod}\quad\omega,
\end{equation} 
with $s=\pm1$, whereas the corresponding eigenstates read as 
\begin{equation}\label{evfull}
|\psi^s_{n}\rangle=\left(
\begin{array}{c}
  f^{s}_ n|n\rangle\\
  sf^{-s}_{ n}|n-2\rangle
\end{array}
\right),
\end{equation}
where  $n=1, 2,\dots$ label the shifted Landau levels. 
We also have defined the coefficients
\begin{equation}
f^s_{n}=\sqrt{\frac{\epsilon_n+s[U- (n-1)\xi]}{2\epsilon_n}},
\end{equation}
which, as expected, reduce to the unperturbed expressions as $\xi\rightarrow0$.
The normalized quasienergies for $n=2\rightarrow5$ are given in  \figurename{2} for the interesting set of parameters $U/\gamma=2\xi/\gamma=0.1$. The upper left 
(right) panel shows the quasi-energy spectrum as a function of normalized cyclotron frequency for the $n=2$ ($n=3$) LL, whereas the lower left (right) panel 
corresponds to $n=4$ ($n=5$). In all panels, the dotted black line gives the static unbiased spectrum, the continuous black (red) curve corresponds to static 
(driven) biased regimes. Interestingly, we notice that the level-dependend bandgap is such that one can realize  configurations where the driven regime is gapless (upper 
right panel), or the driven regime mimics the static biased scenario (lower right panel) which indeed shows the tunability of the photodinduced bandgap $\Delta_n$. 
However, we emphasize that although the photoinduced bandgap might seem to lead to identical physical behaviour of the pseudospin polarization, we will show 
below that this actually not  the case since the interference among the driven eigenstates mixes 
the static eigentstates with different weights. The latter, gives rise to a time-dependent term that is directly proportional to the driving strength. 
Notice however that, for higher LL the photoinduced bandgap continuously grows until the restriction $\Delta_n=\omega$ is reached which is a consequence of the periodicity of the 
quasienergy spectrum.\\ 
\noindent Having dealt with the photoinduced bangdap spectrum we can study the radiation field effects on the layer-dependent pseudospin polarization 
$\tau_z(t)=\langle\Psi(t)|\tau_z|\Psi(t)\rangle$, 
and we can interpret its fluctuations as an indirect measure of the angular momentum exchange among the graphene Dirac fermions and the electromagnetic 
field. That is, it provides information about the photodinduced dynamical hopping  between the upper and lower layer of BLG. In order to gain further physical insight, we first show 
the effects of the radiation field by considering the initial state as an eigenstate (\ref{evec0}) of the static effective two-band Hamiltonian 
$H_2|\phi_{ns}\rangle=E_{ns}|\phi_{ns}\rangle$, with $n\neq0,1$. After somewhat lengthy calculations 
(presented in the appendix) we get
\begin{eqnarray}
\langle\tau_z(t,\xi)\rangle_{ns}&=&\frac{s\Delta_n}{E_n}\Bigg(1+\frac{(n-1)\xi U}{\epsilon_n^2}\Bigg)+\nonumber\\
&&s\Bigg(\frac{\Omega_c^2n(n-1)^2}{E_n\epsilon^2_n}\Bigg)\xi\cos2\epsilon_n t,\quad n\ge2,\nonumber\\
\end{eqnarray}
where $\Delta_n=U-(n-1)\xi$. As expected, in the limit $\xi\rightarrow0$, for which $\Delta_n\rightarrow U$ and $\epsilon_n\rightarrow E_n$, one recovers the result (\ref{polar0}). 
The time average of the polarization in one period $\langle\tau_z\rangle=1/T\int^T_0\langle\tau_z(t,\xi)\rangle dt$ gives
\begin{figure*}\label{fig3}
\includegraphics[height=5cm]{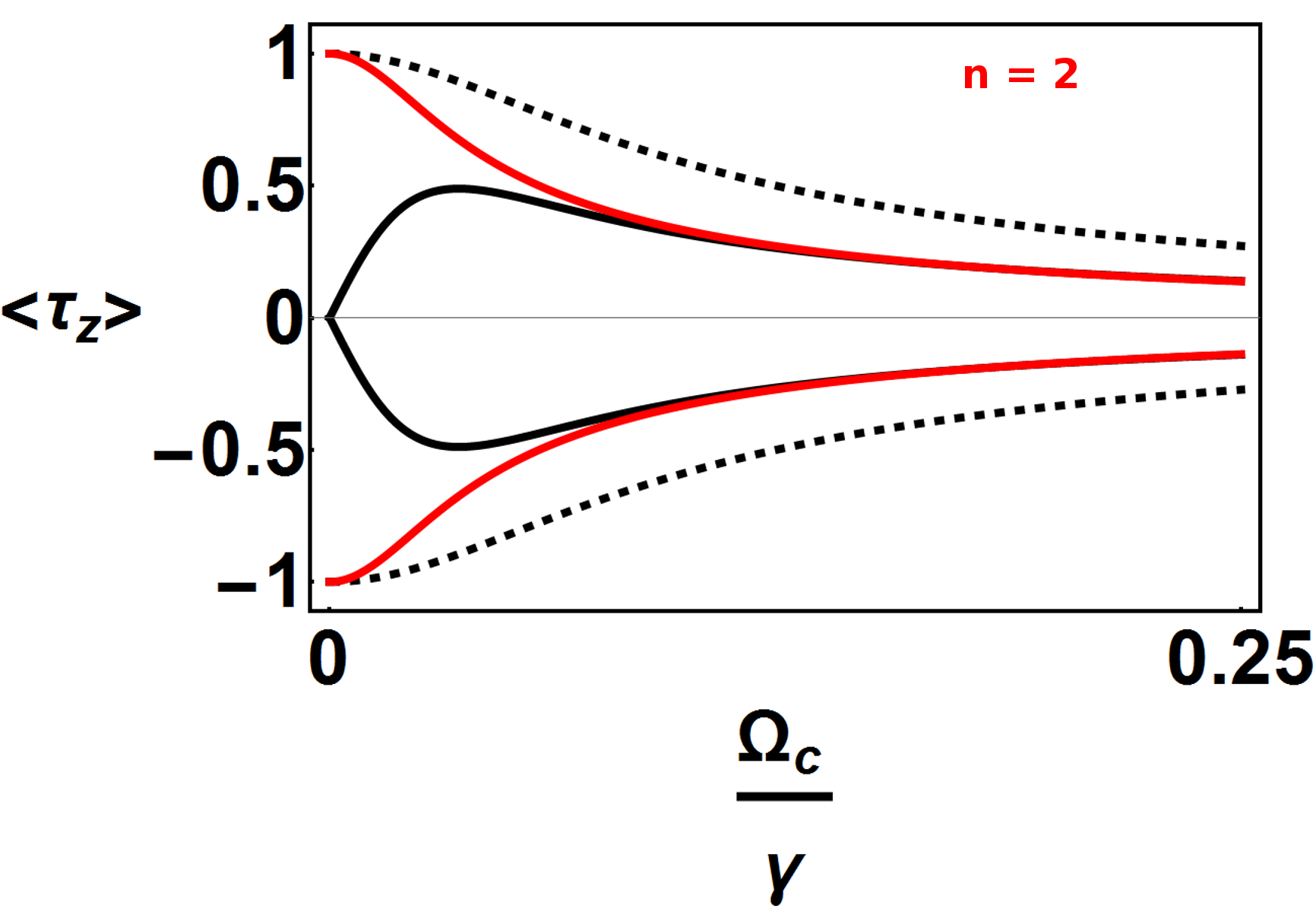}\hspace{.5cm}\includegraphics[height=5cm]{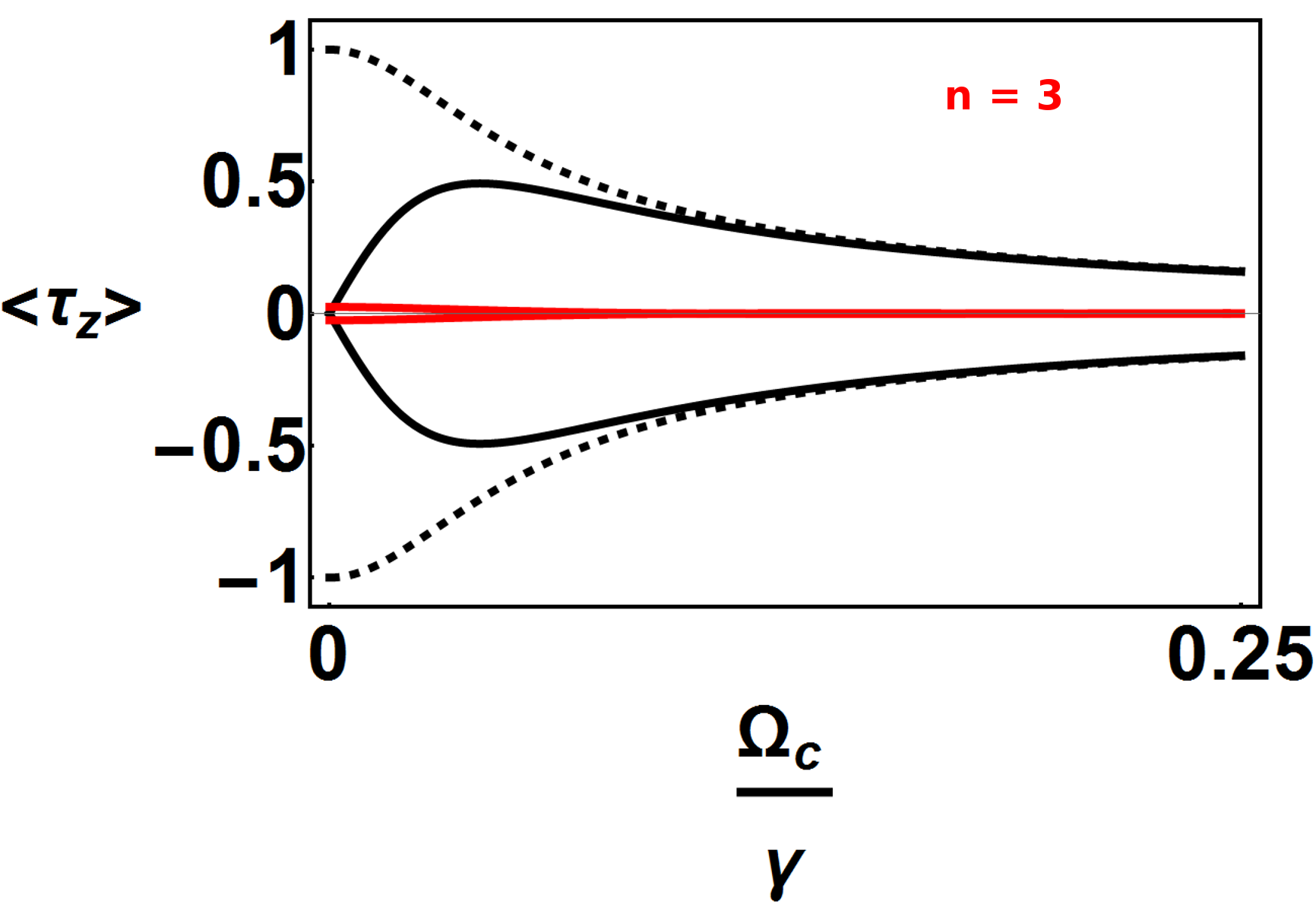}\\
\includegraphics[height=5.cm]{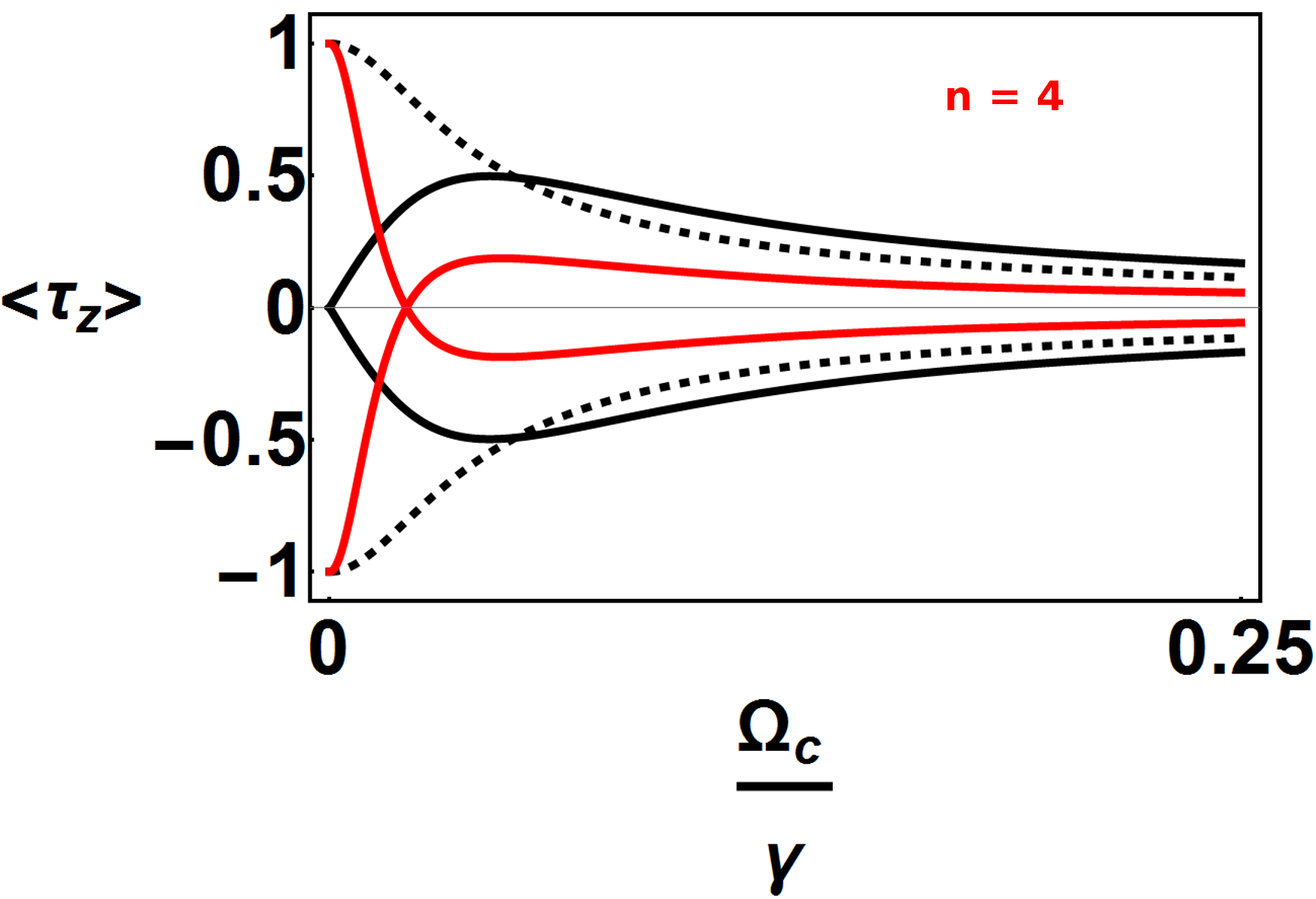}\hspace{.5cm}\includegraphics[height=5.cm]{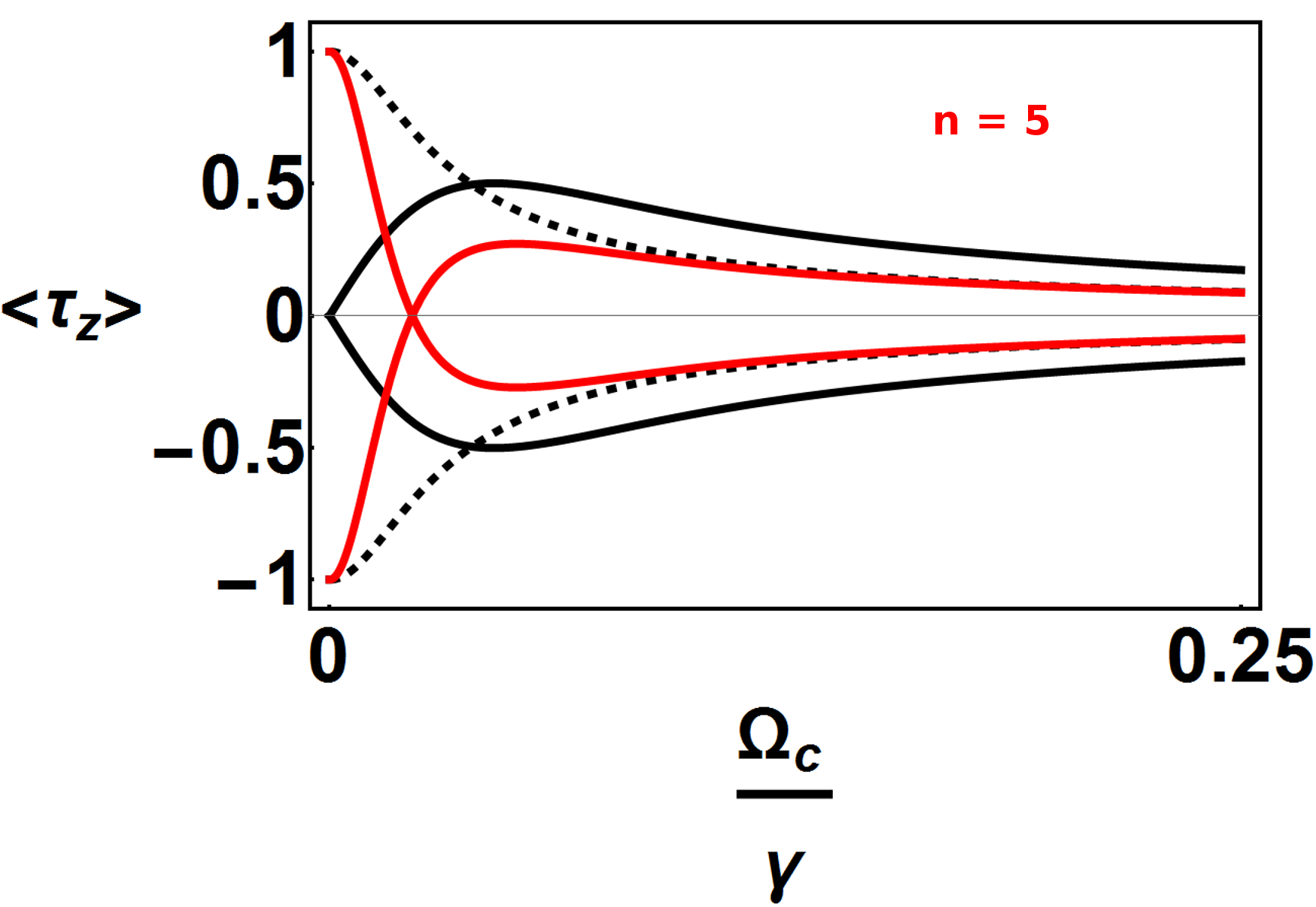}
\caption{(Color online) Averaged pseudospin polarization for $n=2$ (upper left), $n=3$ (upper right), $n=4$    (lower left), $n=5$ (lower right) as functions of normalized cyclotron energy. 
Here, the dashed  lines represent the static scenario for $U=0.1\gamma$, whereas the black (red) corresponds to the unbiased (biased) driven regime for the 
effective coupling value set to $\xi/\gamma=0.05$ (see discussion in the main text).}
\end{figure*}
\begin{eqnarray}
\langle\tau_z\rangle_{ns}&=&\frac{s\Delta_n}{E_n}\Bigg(1+\frac{(n-1)\xi U}{\epsilon_n^2}\Bigg)+\nonumber\\
&&s\Bigg(\frac{\Omega_c^2n(n-1)^2}{E_n\epsilon^2_n}\Bigg)\,\xi\,{\sinc}\, 2\epsilon_n T,\quad n\ge2,\nonumber\\
\end{eqnarray}
\noindent with ${\rm\sinc}\! x=\sin x/x$. The pseudospin polarization is plotted in \figurename{3} for the $n=2\rightarrow5$ LL states. 
We can observe that at low magnetic fields ($\Omega_c\rightarrow0$) the effective driven pseudospin polarization effects can be enhanced within 
the driven scenario when the bias voltage is present. As mentioned before, this can be explained by the interplay of the driving field and this bias 
which provides a LL-dependent bandgap favoring the interlayer hopping and hence the fluctuation in this physical quantity. Indeed, for any finite value of 
the quantizing magnetic field, the unbiased driven polarization shows a LL independent behavior which indicates that addressing the LLs requires the 
presence of the driving field for pseudospin inversion.\\

\noindent The most general scenario can be considered by writing the initial state as a linear superposition of the static eigenstates
\begin{equation}\label{super}
|\Psi(0)\rangle=c_0|\phi_0\rangle+c_1|\phi_1\rangle+\sum_{s=\pm1}\sum_{n\ge2}c_{ns}|\phi_{ns}\rangle
\end{equation} 
where the expansion coefficients satisfy the normalization condition $|c_0|^2+|c_1|^2+\sum_{s=\pm1}\sum_{n\ge2}|c_{ns}|^2=1$, and we have explicitly separated the $n=0,1$ eigenstates since they are degenerate in the pseudospin degree of freedom, as was discussed previously.
Thus, the calculation of the pseudospin polarization gives now
\begin{equation}
\tau_z(t)=|c_0|^2+|c_1|^2+\tilde{\tau}_z(t)
\end{equation}
where the time-dependent contribution reads
\begin{widetext}
\begin{eqnarray}\label{tztilde}
\tilde{\tau}_z(t)&=&\sum_{s=\pm1}\sum_{n\ge2}\frac{1}{\epsilon_n^2E_n}\Big[s|c_{ns}|^2\Big(\Delta_n[U\Delta_n+\Omega_c^2n(n-1)]+\Omega_c^2n(n-1)^2\xi\cos2\epsilon_nt\Big)+\nonumber\\
&&c_{n,-s}^{*}c_{ns}\Omega_c\sqrt{n(n-1)}\Big\{[\Omega_c^2 n(n-1)+U\Delta_n]\cos 2\epsilon_n t-(n-1)\xi\Delta_n-is\epsilon_nE_n\sin2\epsilon_nt\Big\}.
\end{eqnarray}
\end{widetext}
\begin{figure*}\label{fig4}
\includegraphics[height=5.5cm]{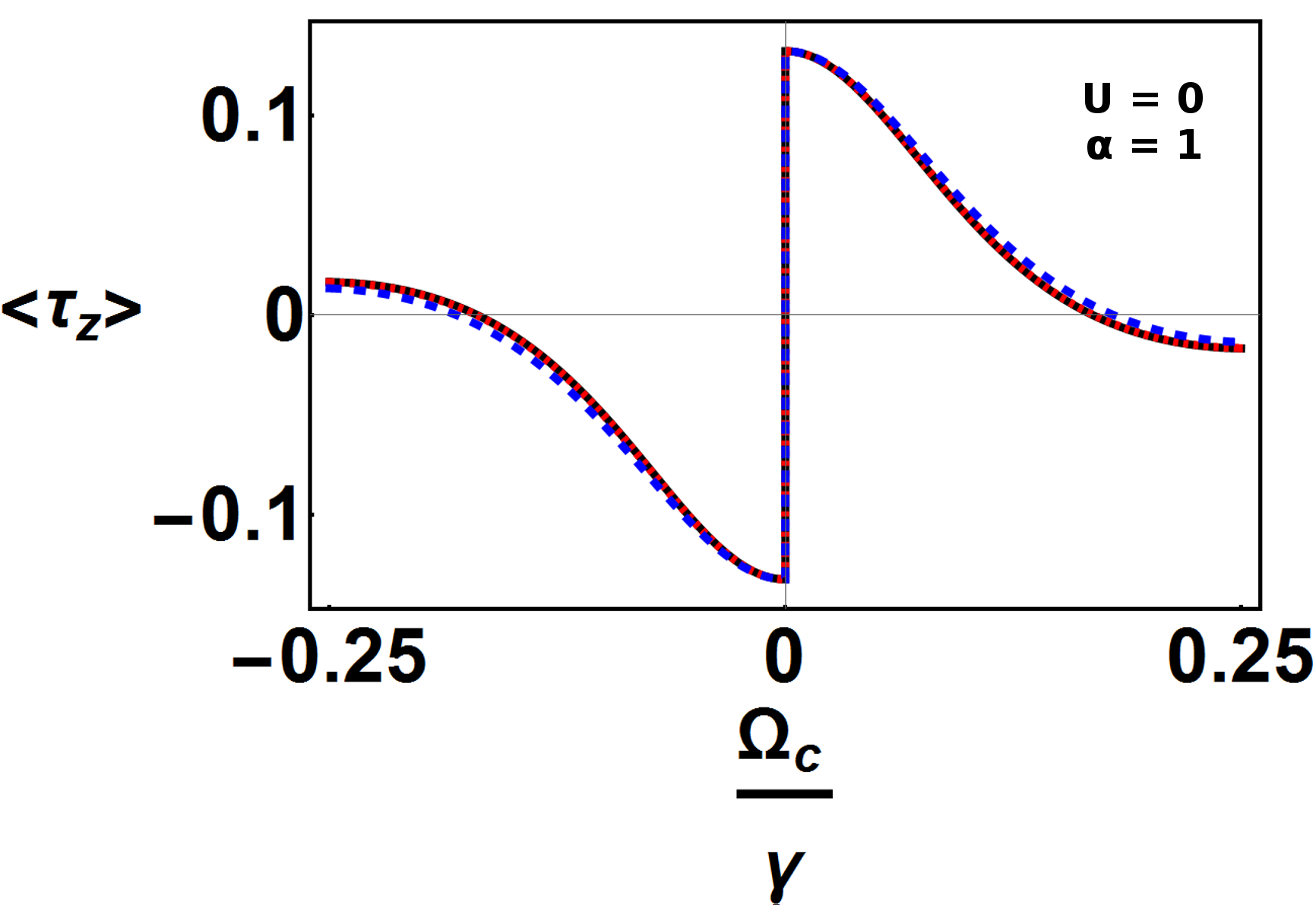}\hspace{.5cm}\includegraphics[height=5.5cm]{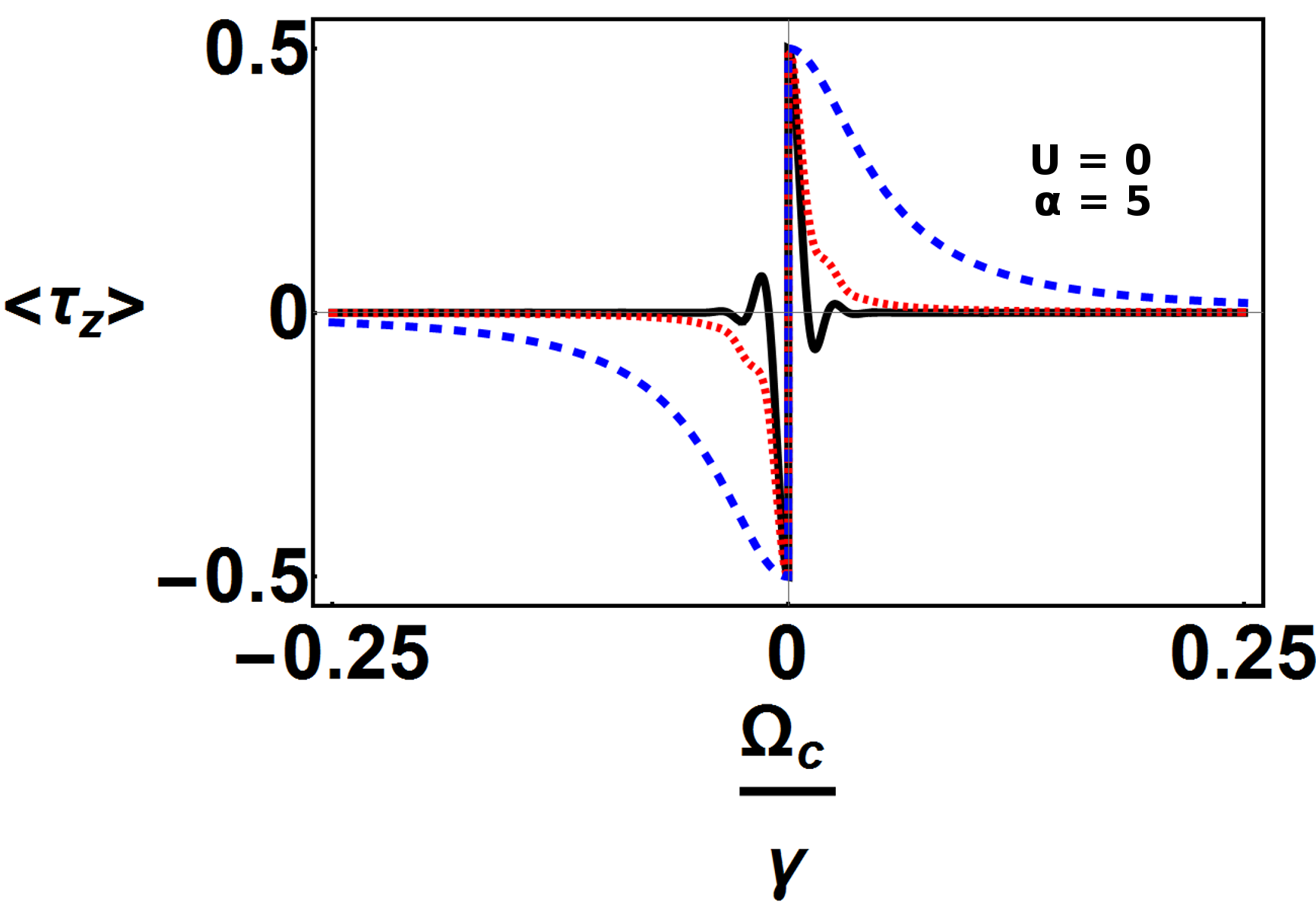}\\
\includegraphics[height=5.5cm]{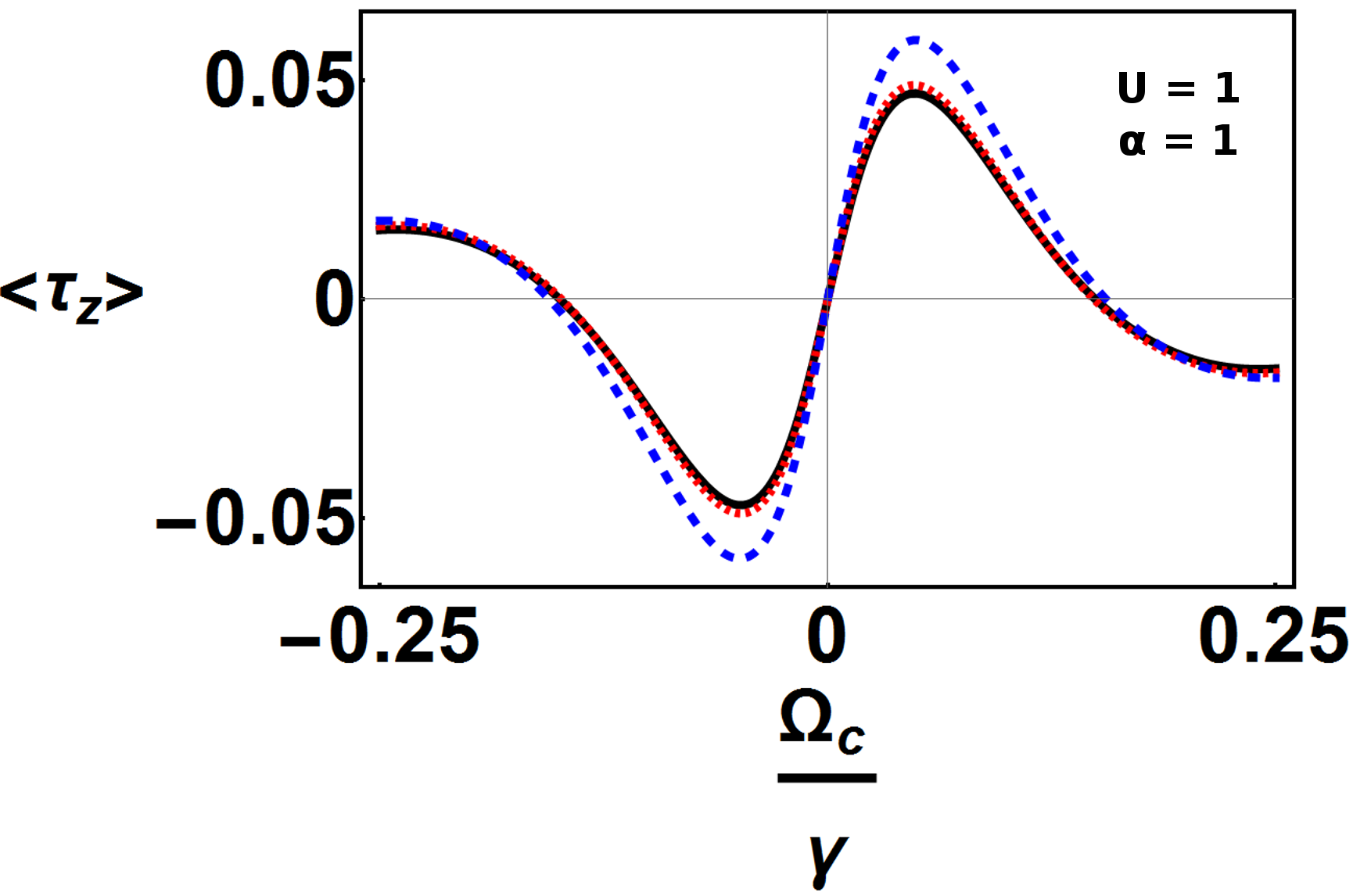}\hspace{.5cm}\includegraphics[height=5.5cm]{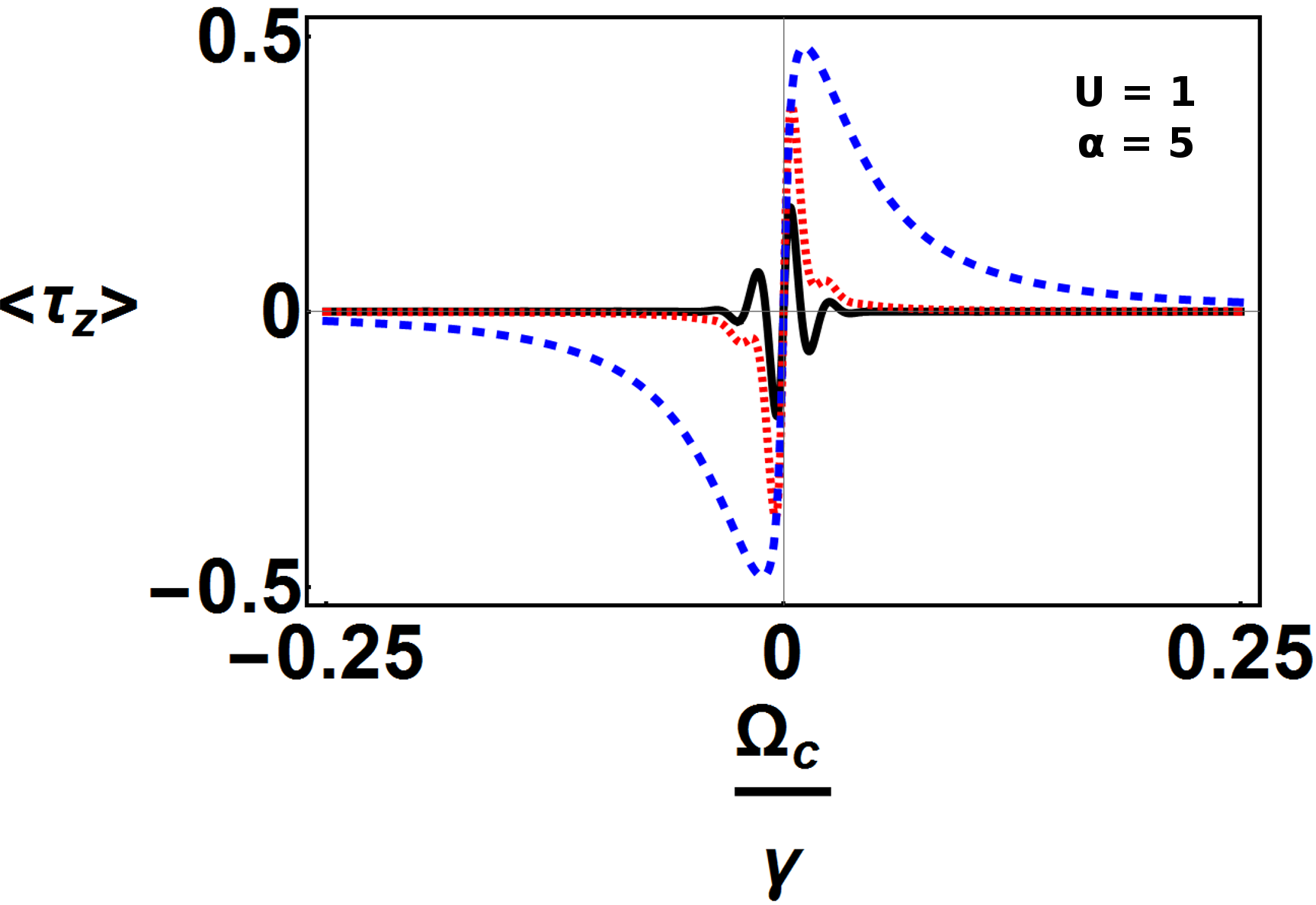}
\caption{(Color online) Averaged pseudospin polarization for the coherent state configuration. The black continuous line represents the static configuration whereas the red (blue) dashed line corresponds to $\xi=0.01\gamma$ ($\xi=0.05\gamma$). 
The upper left (right) panel corresponds to the unbiased case for a coherent state parameter $\alpha=1$ ($\alpha=5$), 
whereas the lower left (right) panel corresponds to the biased $U=0.1\gamma$ scenario for $\alpha=1$ ($\alpha=5$).}
\end{figure*}
As the $n=0,1$ solutions are still eigenstates of $H_{2F}$, there are not pseudospin oscillations in this subspace. Thus, this quantity 
cannot detect any relative phase among these two lower LL states but it does so for the higher LL. Then, we focus our attention on the oscillatory part 
$\tilde{\tau}_z(t)$. For clarity of the analysis we have assumed for simplicity, but without loss of generality, that the expansion coefficients are 
independent of the pseudospin degree of freedom $c_{ns}=c_n/\sqrt{2}$. We notice that under this assumption, the first and last terms in equation 
(\ref{tztilde}) vanish which further simplifies the resulting expression for the polarization. Then the $n\ge2$ LL contributions to the pseudospin polarization
now have the form  
\begin{widetext}
\begin{eqnarray}\label{tztildes}
\tilde{\tau}_z(t)&=&\sum_{n\ge2}\frac{|c_n|^2\Omega_c\sqrt{n(n-1)}}{2E_n}\Bigg\{\cos 2\epsilon_n t-\frac{(n-1)\xi\Delta_n(1-\cos 2\epsilon_n t)}{\epsilon_n^2}\Bigg\},
\end{eqnarray}
\end{widetext}
Within this approximated scenario we notice that the selection of real expansion coefficients allow us a more transparent theoretical description 
of the pseudospin polarization effects. Clearly by selecting $c_0=c_1=0$ implies that only pseudospin oscillations are considered. Upon averaging we get the result 
\begin{widetext}
\begin{eqnarray}\label{tzave}
\langle\tilde{\tau}_z\rangle&=&\sum_{n\ge2}\frac{|c_n|^2\Omega_c\sqrt{n(n-1)}}{2E_n}\Bigg[\sinc\Big(\frac{4\pi\epsilon_n}{\omega}\Big)-\frac{(n-1)\xi\Delta_n[1-\sinc (4\pi\epsilon_n/\omega)]}{\epsilon_n^2}\Bigg].
\end{eqnarray}
\end{widetext}
We have previously considered a coherent state superposition for which $|c_n|^2=e^{-|\alpha|^2}|\alpha|^{2n}/n!$, and show that interesting dynamical 
effects might arise in monolayer graphene LL\cite{39}. Within the formulation of quantum optics, the parameter $\langle n\rangle=|\alpha|^2$ gives a measure of the average 
occupation of the coherent state. 
In order to determine the radiation field effects in a coherent state superposition, we select again 
as initial configuration a coherent state and in  \figurename{4}  we show $\langle\tau_z\rangle$  as a function of $\Omega_c/\gamma$ for different cases. The general
outcome is that for both biased and unbiased regimes, the pseudospin polarization amplitude can be enhanced by means of the radiation field, and it takes typically longer for the driven
pseudospin oscillation to decay. This shows that driving the system by monochromatic radiation affords a better control mechanism to address the pseudospin degree 
of freedom as compared to the role of the bias gate voltage allowing for longer and more pronounced pseudospin polarization effects.
\section{discussion and concluding remarks}
We have analyzed the dynamical modulation of the Landau level structure of biased bilayer graphene subject to circularly polarized terahertz radiation. By means of a perturbative 
semi-analytical treatment  we found that the most salient feature of the photoinduced modulation is to introduce a level dependent bandgap that provides an 
additional control parameter to modulate the electronic properties of the low energy particles present in the bilayer configuration.  
Nontrivial behavior of the pseudospin degree of freedom  can be observed via the oscillations in the  associated polarization dynamics. We show that the
$n\le2$ LL transitions are crucial to obtain a finite polarization for an initially prepared coherent state. In this manner, driving the system by 
monochromatic radiation could afford a better control mechanism to manipulate the pseudospin degree 
of freedom as compared to the sole application of the static bias gate voltage. In addition, we have also shown that in the driven scenario longer and more 
pronounced pseudospin polarization effects can be realized.
Indeed, this is a novel feature of the  driven scenario since the main new  physical features are absent within the subspace spanned by the degenerate $n=0,1$ LL states. We would expect that the reported photoinduced gap modulation and pseudospin oscillations could be detected through the reemitted dipolar radiation 
from the oscillating charge carriers as it was proposed in reference\cite{rusin} or with an appropriate modification of the experimental result reported 
recently\cite{exp-2019} using a pump-probe femtosecond time- and angle-resolved photoemission spectroscopy (tr-ARPES) technique. In such experiment, a laser 
source is used to both map the energies of the excited states as well as 
follow the associated momentum-resolved population dynamics and serves to capture the transient population decay processes. The setup allow them to reach 
the femtosecond time scales associated to $\textrm{In/Si}(111)$. We emphasize  that the frequencies of interest in our model would lie in the near infrared
region. Therefore,  we could expect the detection scheme of our proposal to be in a much lower frequency scale associated to picosecond processes 
as has already been experimentally implemented recently in reference \cite{gedik}.\\
Since in this work we have considered pristine graphene samples, we would like to make a final remark on the role of defects in our results. Experimental evidence shows that defects might appear during the synthesis process of the sample which, at the nanoscale, might lead to interesting new phenomena since they could be exploited to generate novel, innovative and useful materials and devices\cite{42}. For instance, these defects have been observed in situ via transmission electron microscopy\cite{43}. The authors of reference\cite{44}, have reported that point defects lead to notable paramagnetism but no magnetic ordering could be detected down to liquid helium temperatures, whereas the authors of reference\cite{gedik} address the role of topological defects in photoinduced phase transitions. They show that long-range order is inhibited and is only restored when the defects annihilate. They also argue that their results would provide a framework for understanding other photoinduced phase transitions by identifying the generation of defects as a governing mechanism. Thus, considering defects in our setup should lead to further interesting results that would be addressed in future research. A very interesting review on the role of structural defects is given in reference\cite{45}. In addition, we would expect that our results could also pave the road to discussing photoinduced Landau levels in graphene heterostructures with other materials such as black phosphorus\cite{46,47,48} which would be addressed elsewhere. In summary, we have shown that photoinduced enhancement of the pseudospin polarization in AB bilayer graphene can be achieved within experimentally accessible parameter regimes. We expect that our results could lead to further interesting physical scenarios in other two-dimensional materials as black phosphorus or transition metal dichalcogenides, among others.
 
{\it{Acknowledgments}--} 
AL acknowledges useful discussions with Ernesto Medina and Leonardo Basile. This work has been supported by CEDIA via the project CEPRA-XII-2018-06 
``Espectroscop\'ia Mec\'anica: Transporte interacci\'on materia radiaci\'on''. AL thanks the University of Lorraine for partial financial support through research visits via CNRS-(PICS) project. AL and FM acknowledge funding from the project DGPA-PAPIIT IN111317.
\section{appendix:  Derivation of the pseudospin polarization}
Here we present some algebraic steps leading to the expressions of the photoinduced polarization effects. First, we give the results for an initially prepared eigenstate of the static effective two-band Hamiltonian
\begin{align}
\ket{\phins} = \qty(\begin{array}{l}
                      \bpsn \ketn \\
                      \stau\bmsn \ket{n-2} 
                    \end{array}),
\end{align}
where we have defined the coefficients 
\begin{align}\label{bn}
\bpsn = \qty(\frac{E_n + \stau U}{2E_n})^{1/2},
\end{align}
whereas, the approximate Floquet states are given as
\begin{align}
\ket{\psi_{m s}} = \qty(\begin{array}{l}
                       f^{s}_m \ket{m} \\
                     sf^{-s}_m \ket{m-2} 
                    \end{array}),
\end{align}
with corresponding coefficients
\begin{align}\label{fn}
f^{s}_m= \qty(\frac{\varepsilon_m + s \Delta_m}{2\varepsilon_m})^{1/2}.
\end{align}
Then the polarization calculation is as follows
\begin{widetext}
\begin{align}
 \bstns &= \mel{\phins}{e^{i H_{2F}t}\stau_z e^{-i H_{2F}t}}{\phins},\\
 &= \sssp\smmp \bra{\phins}\ket{\psi_{m^\prime s^\prime}}
    \mel{\psi_{m^\prime s^\prime}}{\stau_z}{\psi_{ms}}
    \bra{\psi_{ms}}\ket{\phins}e^{-\qty(\varepsilon_{ms}-\varepsilon_{m^{\prime}s})t},\\
 &= \sssp\smmp \bra{\phins}\ket{\psi_{m s^\prime}}
    \bra{\psi_{ms}}\ket{\phins}
    \qty(f^{s^\prime}_m f^s_m - f^{-s}_m f^{-s^{\prime}}_m s s^{\prime})
    e^{-i\qty(s-s^\prime)\varepsilon_m t}\delta_{mn}\delta_{m^\prime n},\\
 &= \sum_{s=\pm\tau} \qty[\qty(b^{\tau}_n f^s_n + \stau s b^{-\tau}_n f^{-s}_n)^2 \qty[(f^s_n)^2 - (f^{-s}_n)^2] 
  + 2\qty( \bpsn f^{-s}_n - \stau s \bmsn f^s_n) \qty(f^s_n \bpsn + s \stau f^{-s}_n \bmsn)
  \qty(f^s_n f^{-s}_n) e^{-2is\varepsilon_n t} ]
\end{align}
\end{widetext}
Using the definitions (\ref{bn}) and (\ref{fn}), we get
\begin{align}
\bstns=&\frac{\tau\Dn}{\varepsilon_n E_n}\Bigg(\frac{U\Dn}{\varepsilon_n}
+\frac{\Omega_c^2 n(n-1)}{E_n}\Bigg)\nonumber\\
&+\frac{\tau\Omega_c^2n(n-1)^2\xi}{E_n\varepsilon_n^2}\cos2\varepsilon_n t.
\end{align}

For the general scenario, the initial state is given as a superposition of the static Hamiltonian eigenstates
\begin{equation}
\ket{\Psi(0)}=c_0\ket{\phi_0}+c_1\ket{\phi_1}+\sum_{ns}c_{ns}\ket{\phi_{ns}},
\end{equation}
where, as it is discussed in the main text, we have isolated the $n=0,1$ LLs which are occupied in a single subspace of the pseudospin degree of freedom. 
In this expressions, the expansion coefficients satisfy the normalization condition $|c_0|^2+|c_1|^2+\sum_{ns}|c_{ns}|^2=1$. Now, taking into account that 
the $n=0,1$ LL remain as eigenstates of the Floquet Hamiltonian $H_{2F}$, we get for the pseudospin polarization dynamics 
$\langle\tau_z(t)\rangle=\langle\Psi(0)|e^{iH_{2F}t}\tau_ze^{-iH_{2F}t}|\Psi(0)\rangle=|c_0|^2+|c_1|^2+\langle\tilde{\tau}_z(t)\rangle$.  
The time-dependent part is a bit lengthy but can be explicitly worked out  as follows
\begin{widetext}
\begin{align}
 \langle\tilde{\tau}_z(t)\rangle &=\sum_{n\tau}\sum_{n^{\prime}\tau^{\prime}} c^{*}_{n\tau}c_{n^{\prime}\tau^{\prime}} 
 \mel{\phins}{e^{i H_{2F}t}\stau_z e^{-i H_{2F}t}}{\phi_{n^{\prime}\tau^{\prime}}},\\
 &=  \sum_{n\tau}\sum_{n^{\prime}\tau^{\prime}} c^{*}_{n\tau}c_{n^{\prime}\tau^{\prime}}\sssp\smmp \bra{\phins}\ket{\psi_{m^\prime s^\prime}}
    \mel{\psi_{m^\prime s^\prime}}{\stau_z}{\psi_{ms}}
    \bra{\psi_{ms}}\ket{\phi_{n^{\prime}\tau^{\prime}}}e^{-\qty(\varepsilon_{ms}-\varepsilon_{m^{\prime}s})t},\\
 &=\sum_{n\tau}\sum_{n^{\prime}\tau^{\prime}} c^{*}_{n\tau}c_{n^{\prime}\tau^{\prime}} \sssp\smmp \bra{\phins}\ket{\psi_{m s^\prime}}
    \bra{\psi_{ms}}\ket{\phi_{n^{\prime}\tau^{\prime}}}
    \qty(f^{s^\prime}_m f^s_m - f^{-s^{\prime}}_m f^{-s}_m s s^{\prime})
    e^{-i\qty(s-s^\prime)\varepsilon_m t}\delta_{mn}\delta_{m^\prime n},\\
 &=\sum_{n}\sum_{\tau\tau^{\prime}} c^{*}_{n\tau}c_{n\tau^{\prime}} \sum_{s=\pm\tau^{\prime}}\qty{\bra{\phi_{n\tau}}\ket{\psi_{ns}}\bra{\psi_{ns}}\ket{\phi_{n\tau^{\prime}}} \qty[(f^s_n)^2 - (f^{-s}_n)^2] 
  + 2\bra{\phi_{n\tau}}\ket{\psi_{n,-s}}\bra{\psi_{ns}}\ket{\phi_{n\tau^{\prime}}}
  \qty(f^s_n f^{-s}_n) e^{-2is\varepsilon_n t} ]}\\
  &=\sum_{n\tau}|c_{n\tau}|^2\qty{\qty{|\bra{\phi_{n\tau}}\ket{\psi_{n\tau}}|^2-|\bra{\phi_{n\tau}}\ket{\psi_{n,-\tau}}|^2}\qty[(f^{\tau}_n)^2 - (f^{-\tau}_n)^2] 
  + 4f^{\tau}_n f^{-\tau}_n\Re\qty{\bra{\phi_{n\tau}}\ket{\psi_{n,-\tau}}\bra{\psi_{n,\tau}}\ket{\phi_{n\tau}}
   e^{-2i\tau\varepsilon_n t}}}\nonumber\\
&-\sum_{n\tau} c^{*}_{n\tau}c_{n,-\tau}\qty{\bra{\phi_{n\tau}}\ket{\psi_{n,-\tau}}\bra{\psi_{n,-\tau}}\ket{\phi_{n,-\tau}}-\bra{\phi_{n\tau}}\ket{\psi_{n\tau}}\bra{\psi_{n\tau}}\ket{\phi_{n,-\tau}}}\qty[(f^{\tau}_n)^2 - (f^{-\tau}_n)^2]\nonumber\\ 
  &+\sum_{n\tau} c^{*}_{n\tau}c_{n,-\tau} 4f^{\tau}_n f^{-\tau}_n\Re\qty{\bra{\phi_{n\tau}}\ket{\psi_{n\tau}}\bra{\psi_{n,-\tau}}\ket{\phi_{n\tau}}
   e^{2i\tau\varepsilon_n t}}.
\end{align} 
\end{widetext}
Up to this point, no assumption has been made about the expansion coefficients. To further simplify the previous expression we consider the experimentally relevant situation for which these parameters are pseudospin-independent, i.e. $c_{n\tau}=c_n/\sqrt{2}$. Within this regime, upon substitution of the dot products, we get the compact expression
\begin{widetext}
\begin{align}
 \langle\tilde{\tau}_z(t)\rangle &=\sum_{n}|c_{n}^2|\sum_\tau\Big[|\bra{\phi_{n\tau}}\ket{\psi_{n\tau}}|^2-|\bra{\phi_{n\tau}}\ket{\psi_{n,-\tau}}|^2\nonumber\\
 &-(\bra{\phi_{n\tau}}\ket{\psi_{n,-\tau}}\bra{\psi_{n,-\tau}}\ket{\phi_{n,-\tau}}-\bra{\phi_{n\tau}}\ket{\psi_{n\tau}}\bra{\psi_{n\tau}}\ket{\phi_{n,-\tau}})\Big]\qty[(f^{\tau}_n)^2 - (f^{-\tau}_n)^2]\nonumber\\
 &+4\sum_{n}|c_{n}^2|\sum_\tau f^{\tau}_n f^{-\tau}_n\Re\qty{\bra{\phi_{n\tau}}\ket{\psi_{n,-\tau}}\bra{\psi_{n,\tau}}\ket{\phi_{n\tau}}
   e^{-2i\tau\varepsilon_n t}}.
\end{align}
\end{widetext}
Upon substitution of the matrix elements 
$\bra{\phi_{n,\tau}}\ket{\psi_{n',\tau'}}$, and using equations (\ref{bn}) and (\ref{fn}) we arrive at the result
given in equation (\ref{tztilde}). It is important to mention that in order to obtain the reported results we are assuming that, to leading order in the 
parameter $\lambda=\xi/\omega_c$, 
the relation $\bra{m}\ket{n}=\delta_{nm}$ among
the original number operator eigenstates $a^\dagger a\ket{n}=n\ket{n}$ and the shifted ones 
$b^\dagger b\ket{m}=m\ket{m}$ is valid. Indeed, we expect that the scenario described in this work should hold in typical experimental setups in which 
the higher order corrections are negligible whenever $\xi\ll\Omega_c\ll\gamma$.

\end{document}